\documentclass[a4paper]{jpconf}
\usepackage{graphicx}
\newcommand{\ba}{\begin{eqnarray}}
\newcommand{\ea}{\end{eqnarray}}

\begin{document}

\title{Odd-mass nuclei in the cluster shell model}

\author{R. Bijker and A.H. Santana-Vald\'es}
\address{Instituto de Ciencias Nucleares, 
Universidad Nacional Aut\'onoma de M\'exico, 
A.P. 70-543, 04510 Ciudad de M\'exico, M\'exico}
\ead{bijker@nucleares.unam.mx}

\begin{abstract}
In this contribution, we present the cluster shell model which is analogous to the Nilsson model, 
but for cluster potentials. Special attention is paid to the consequences of the discrete symmetries 
of three $\alpha$-particles in an equilateral triangle configuration. This configuration is characterized 
by a special structure of the rotational bands which can be used as a fingerprint of the underlying 
geometric configuration. The cluster shell model is applied to the nucleus $^{13}$C. 
\end{abstract}

\section{Introduction}

Cluster degrees of freedom are very important for the description of light nuclei, in particular 
$k\alpha$ and $k\alpha+x$ nuclei, due to the large binding energy of the $^4$He nucleus. 
Early work on $\alpha$-cluster models goes back to the 1930's with 
studies by Wheeler \cite{wheeler}, and Hafstad and Teller \cite{Teller}, followed by later work 
by Brink \cite{Brink1,Brink2} and Robson \cite{Robson1,Robson2}. Recently, there has been a lot of 
renewed interest in the structure of $\alpha$-cluster nuclei, especially for the nucleus $^{12}$C 
\cite{FreerFynbo}. The measurement of new rotational excitations of the ground state \cite{Freer2007,Kirsebom,Marin} 
and of the Hoyle state \cite{Itoh,Freer2012,Gai,Freer2011} has stimulated a large theoretical effort to 
understand the structure of $^{12}$C (for a review see {\it e.g.} Refs.~\cite{FreerFynbo,Schuck,Freer}). 

In this contribution, we present a study of cluster states in $^{12}$C and $^{13}$C 
in the framework of the algebraic cluster model and the cluster shell model, respectively. 

\section{The algebraic cluster model}

The algebraic cluster model (ACM) is an interacting boson model to describe the relative motion 
of cluster systems (see {\it e.g.} \cite{Bijker2016,PPNP}). The relevant degrees of freedom of a system 
of $k$-body clusters are given by the $k-1$ relative Jacobi coordinates and their conjugate momenta. 
The building blocks of the ACM consist of a vector boson for each relative coordinate and a scalar boson.  
Cluster states are then described in terms of a system of $N$ interacting bosons with angular 
momentum and parity $L^P=1^-$ (vector bosons) and $L^P=0^+$ (scalar bosons). The $3(k-3)$ components 
of the vector bosons together with the scalar boson span a $(3k-2)$-dimensional space with group 
structure $U(3k-2)$. The many-body states are classified according to the totally 
symmetric irreducible representation $[N]$ of $U(3k-2)$. The ACM has a very rich symmetry structure. 
In addition to continuous symmetries like the angular momentum, in case of $\alpha$-cluster nuclei the 
Hamiltonian has to be invariant under the permuation of the $k$ identical $\alpha$ particles. 
Since one does not consider the excitations of the $\alpha$ particles themselves,  
the allowed cluster states have to be symmetric under the permutation group $S_k$.  

The potential energy surface corresponding to the $S_k$ invariant ACM Hamiltonian gives rise to several 
possible equilibrium shapes. The ones of special interest for applications to $\alpha$-cluster nuclei, 
correspond to the geometrical configuration of a dumbbell for $^{8}$Be ($k=2$ with $Z_2$ symmetry), an 
equilateral triangle for $^{12}$C ($k=3$ with $D_{3h} \supset D_3$ symmetry) and a regular tetrahedron 
for $^{16}$O ($k=4$ with $T_d$ symmetry), see Table~\ref{ACM}. Even though these cases do not 
correspond to dynamical symmetries of the ACM Hamiltonian, one can still obtain approximate 
solutions for the rotation-vibration spectrum and electromagnetic properties 
\cite{Valeria,Bijker2002,Bijker2017}.

\begin{table}
\centering
\caption{ACM for $k$-body clusters.}
\label{ACM}
\vspace{10pt}
\begin{tabular}{cccccc}
\hline
\noalign{\smallskip}
$k$ & $U(3k-2)$ & Symmetry & Geometry & Nucleus & Ref \\
\noalign{\smallskip}
\hline
\noalign{\smallskip}
$2$ & $U(4)$  & $Z_2 \sim S_2$ & Dumbbell & $^{8}$Be & \cite{Valeria} \\
$3$ & $U(7)$  & $D_{3h} \supset D_3 \sim S_3$ & Equilateral triangle & $^{12}$C & \cite{Bijker2000,Bijker2002} \\
$4$ & $U(10)$ & $T_d \sim S_4$ & Regular tetrahedron & $^{16}$O & \cite{Bijker2014,Bijker2017} \\
\noalign{\smallskip}
\hline
\end{tabular}
\end{table}

\begin{figure}[b]
\centering
\setlength{\unitlength}{0.7pt} 
\begin{picture}(240,270)(0,0)
\thinlines
\put (  0,  0) {\line(1,0){240}}
\put (  0,270) {\line(1,0){240}}
\put (  0,  0) {\line(0,1){270}}
\put (240,  0) {\line(0,1){270}}
\thicklines
\put ( 30, 60) {\line(1,0){20}}
\put ( 30, 78) {\line(1,0){20}}
\put ( 30, 96) {\line(1,0){20}}
\put ( 30,120) {\line(1,0){20}}
\put ( 30,150) {\line(1,0){20}}
\multiput ( 80, 60)(5,0){23}{\circle*{0.1}}
\thinlines
\put ( 25, 25) {$(00)A$}
\put ( 55, 57) {$0^+$}
\put ( 55, 75) {$2^+$}
\put ( 55, 93) {$3^-$}
\put ( 55,117) {$4^{\pm}$}
\put ( 55,147) {$5^-$}
\thicklines
\put (100, 90) {\line(1,0){20}}
\put (100,108) {\line(1,0){20}}
\put (100,126) {\line(1,0){20}}
\put (100,150) {\line(1,0){20}}
\put (100,180) {\line(1,0){20}}
\multiput (110, 60)(0,5){6}{\circle*{0.1}}
\thinlines
\put ( 95, 25) {$(10)A$}
\put (125, 87) {$0^+$}
\put (125,105) {$2^+$}
\put (125,123) {$3^-$}
\put (125,147) {$4^{\pm}$}
\put (125,177) {$5^-$}
\thicklines
\put (170,126) {\line(1,0){20}}
\put (170,138) {\line(1,0){20}}
\put (170,156) {\line(1,0){20}}
\put (170,180) {\line(1,0){20}}
\put (170,210) {\line(1,0){20}}
\multiput (180, 60)(0,5){13}{\circle*{0.1}}
\thinlines
\put (165, 25) {$(01)E$}
\put (195,123) {$1^-$}
\put (195,135) {$2^{\mp}$}
\put (195,153) {$3^{\mp}$}
\put (195,177) {$4^{\mp+}$}
\put (195,207) {$5^{\mp\pm}$}
\thicklines
\put(25,200) {\circle*{10}} 
\put(65,200) {\circle*{10}}
\put(45,240) {\circle*{10}}
\put(25,200) {\line( 1,0){40}}
\put(25,200) {\line( 1,2){20}}
\put(65,200) {\line(-1,2){20}}
\end{picture}
\caption[Schematic spectrum of a triangular configuration]
{Schematic spectrum of a triangular configuration. The rotational bands are labeled by 
$(v_1 v_2)$ and $t$ (bottom). All states are symmetric under $S_3 \sim D_3$.} 
\label{triangle}
\end{figure}
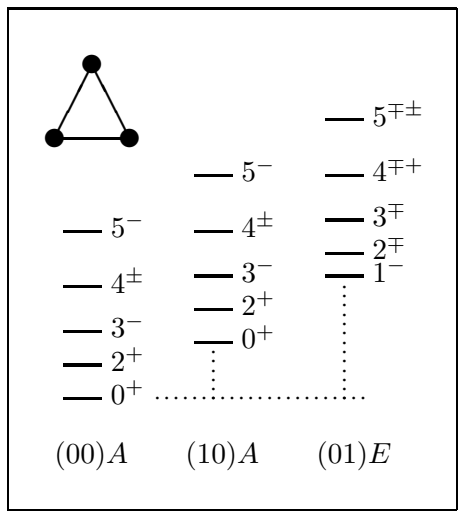

\begin{figure}
\centering
\includegraphics[width=4in]{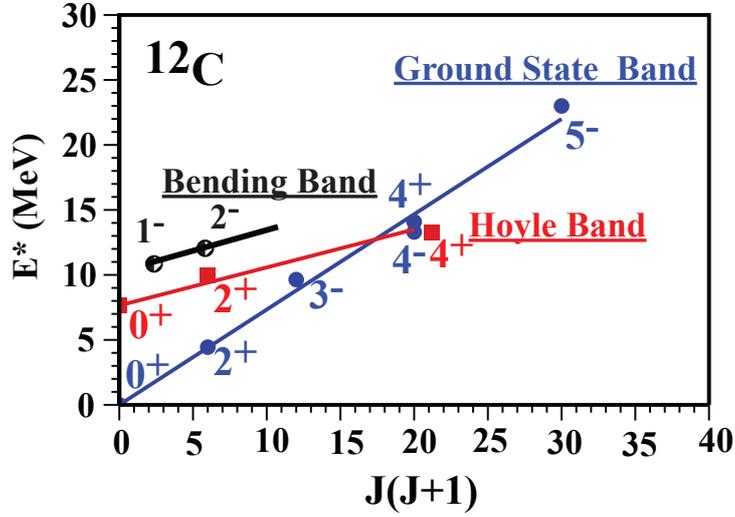} 
\caption[Rotational bands in $^{12}$C]
{Rotational bands in $^{12}$C \cite{Marin}.}
\label{bands3}
\end{figure}

In this contribution, we present the results for three-body clusters with a triangular configuration. 
The rotation-vibration spectrum is given by 
\ba
E \;=\; \omega_{1}(v_{1}+\frac{1}{2}) + \omega_{2}(v_{2}+1) + \kappa \, L(L+1) ~.
\label{energy} 
\ea
The rotational structure of the ground-state band, $(v_1,v_2)=(0,0)$, and the fundamental vibrations, 
$(1,0)$ and $(0,1)$, depends on the $D_{3h}$ point group symmetry of the equilateral triangle 
configuration and is summarized in Fig.~\ref{triangle}. 
The ground-state band is characterized by a rotational sequence involving both positive and 
negative parity states, $L^P=0^+$, $2^+$, $3^-$, $4^{\pm}$, $5^-$, $\ldots$, 
all of which have been observed in $^{12}$C. The so-called Hoyle band has the same structure, 
but so far only the positive parity states have been observed. 
The rotational bands in $^{12}$C are shown in Fig.~\ref{bands3}. 

The matter and charge distributions are usually taken as a Gaussian distribution 
\ba
\rho(\vec{r}) &=& \left( \frac{\alpha}{\pi} \right)^{3/2} 
\sum_{i=1}^{3} \exp \left[ -\alpha \left( \vec{r}-\vec{r}_{i} \right)^{2}\right] ~,
\label{rhor}
\ea
where the $\alpha$-particles are located at a distance $\beta$ from the center of mass with 
$\vec{r_i}=(\beta,\theta_i,\phi_i)$. The charge density is obtained by multiplying Eq.~(\ref{rhor}) 
by $Ze/3$ and the matter density by $Am/3$. 
Electromagnetic transition probabilities can be obtained from the transition form factors, 
which are the matrix elements of the Fourier transform of the charge distribution. 
For transitions along the ground-state band the transition form factors are given in terms of a 
product of a spherical Bessel function and an exponential factor arising from a Gaussian distribution 
of the electric charges \cite{Bijker2002} 
\ba
F(0^+ \rightarrow L^P;q) \;=\; c_L \, j_L(q \beta) \, \mbox{e}^{-q^{2}/4\alpha} ~.
\ea 
The transition form factors depend on the parameters $\alpha$ and $\beta$. 
The value of $\alpha$ is determined from the radius of the $\alpha$-particle 
to be $\alpha=0.56$ fm$^{-2}$ \cite{Sick}, and the value of $\beta$ from the first minimum 
of the elastic form factor of $^{12}$C to be $\beta=1.74$ fm \cite{Bijker2002}.

The transition probabilities $B(EL)$ along the ground-state band can be extracted 
from the form factors in the long wavelength limit 
\ba
B(EL;0^+ \rightarrow L^P) \;=\; \frac{(Ze\beta^{L}c_L)^2}{4\pi} ~, 
\label{BEL}
\ea
with
\ba
c_L^2 \;=\; \frac{2L+1}{3} \left[ 1+2P_{L}(-\frac{1}{2}) \right] 
\ea
and the charge radius from the slope of the elastic form factor in the origin
\ba
\langle r^{2} \rangle^{1/2} \;=\; \sqrt{\beta^{2}+3/2\alpha} ~.
\ea
Eq.~(\ref{BEL}) shows that all electric transitions are related to one another, 
and only depend on the value of $\beta$. For example, quadrupole, octupole and 
hexadecupole transitions are related by
\ba
\frac{B(E3;3_{1}^{-} \rightarrow 0_{1}^{+})}{B(E2;2_{1}^{+} \rightarrow 0_{1}^{+})} 
&=& \frac{5}{2}  \beta^2 ~,
\nonumber\\ 
\frac{B(E4;4_{1}^{+} \rightarrow 0_{1}^{+})}{B(E2;2_{1}^{+} \rightarrow 0_{1}^{+})} 
&=& \frac{9}{16} \beta^4 ~.
\ea
The quadrupole moment can be calculated as \cite{Yale2018}
\ba
Q_{2^+} \;=\; \frac{2}{7} Ze \beta^2 ~, 
\ea
a positive value, as is to be expected for a planar configuration of three $\alpha$-particles. 

Table~\ref{BELC12} shows a good agreement with the experimental values of $^{12}$C which provides 
evidence that the $2_{1}^{+}$ and $3_{1}^{-}$ states belong to the ground-state rotational band of 
the equilateral triangle \cite{Bijker2002}. It is important to note that the ACM results only depend 
on the value of $\alpha$ and $\beta$ which have been determined from other observables.  

\begin{table}[h]
\centering
\caption{$B(EL)$ values, quadrupole moment and charge radius for $^{12}$C. 
Experimental data are taken from \cite{kelley,kumar,reuter}, and the ACM values 
from \cite{Bijker2002}.}
\label{BELC12}
\vspace{10pt}
\begin{tabular}{cccl}
\noalign{\smallskip}
\hline
\noalign{\smallskip}
& ACM & Exp & \\
\noalign{\smallskip}
\hline
\noalign{\smallskip}
$B(E2;2_{1}^{+}\rightarrow 0_{1}^{+})$  & 8.4 & $7.61 \pm 0.42$ & $e^{2}\mbox{fm}^{4}$ \\
\noalign{\smallskip}
$B(E3;3_{1}^{-}\rightarrow 0_{1}^{+})$ & 73 & $104 \pm 14$ & $e^{2}\mbox{fm}^{6}$ \\
\noalign{\smallskip}
$B(E4;4_{1}^{+}\rightarrow 0_{1}^{+})$ & 44 & & $e^{2}\mbox{fm}^{8}$ \\  
\noalign{\smallskip}
\hline
\noalign{\smallskip}
$Q_{2_{1}^{+}}$ & $5.2$ & $5.3 \pm 4.4$ & $e\mbox{fm}^{2}$ \\
\noalign{\smallskip}
$\langle r^{2} \rangle^{1/2}_{\rm ch}$ & $2.389$ & $2.468 \pm 0.012$ & fm \\
\noalign{\smallskip}
\hline
\end{tabular}
\end{table}

\section{The cluster shell model}

\begin{figure}
\centering
\includegraphics[scale=0.6]{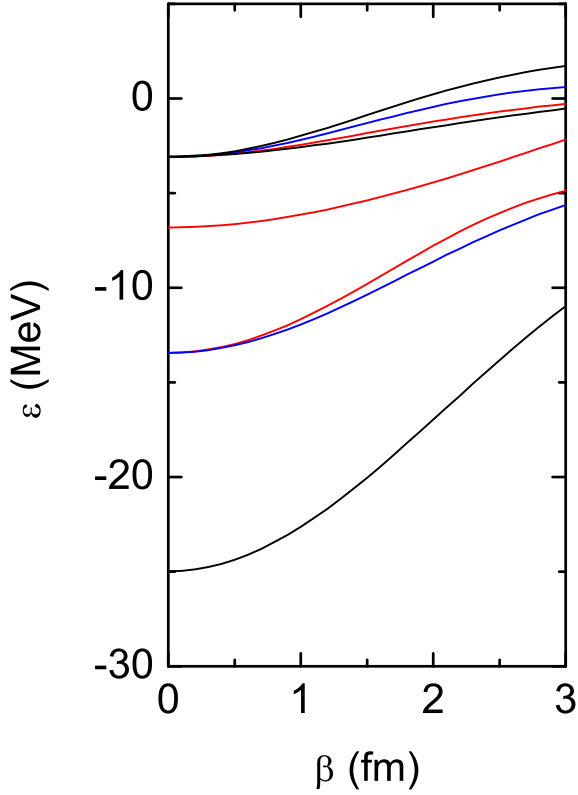}
\caption{Single-particle energies in a cluster potential with ${\cal D'}_{3h}$ triangular symmetry. 
The single-particle levels are labeled by $E_{1/2}^{(+)}$ (black), $E_{1/2}^{(-)}$ (red) and $E_{3/2}$ 
(blue). For $\beta=0$ the ordering of the single-particle orbits is $1s_{1/2}$, $1p_{3/2}$, $1p_{1/2}$ 
and (almost degenerate) $1d_{5/2}$, $2s_{1/2}$.}
\label{splevels}
\end{figure}

For the description of odd-cluster nuclei, the cluster shell model (CSM) was developed recently 
\cite{PPNP,Valeria,CSM}. 
The CSM combines cluster and single-particle degrees of freedom, and is very similar in spirit as the 
Nilsson model \cite{Nilsson}, but in the CSM the odd nucleon moves in the deformed field generated by the 
(collective) cluster degrees of freedom. The Hamiltonian is written as
\ba
H \;=\; T + V(\vec{r}) + V_{\rm so}(\vec{r}) + V_{\rm C}(\vec{r}) ~,
\ea
{\it i.e.} the sum of the kinetic energy, a central potential obtained by convoluting 
the density $\rho(\vec{r})$  of Eq.~(\ref{rhor}) with the interaction between the $\alpha$-particle 
and the nucleon, a spin-orbit interaction and, for an odd proton, a Coulomb potential \cite{CSM}. 
Fig.~\ref{splevels} shows the splitting of single-particles levels of a neutron moving in the deformed 
potential generated by a triangular configuration of three $\alpha$-particles as a function of $\beta$. 
The solutions of the CSM Hamiltonian, $\chi_{\Omega}$ with energy $\epsilon_{\Omega}$, are labeled by the 
irreducible representations of the double group $D'_{3h}$: $\Omega=E_{1/2}^{(-)}$, $E_{1/2}^{(+)}$ and 
$E_{3/2}$, each of which is doubly degenerate. The resolution of single-particle levels into representations 
of $D'_{3h}$ is shown in Table~\ref{splitting}. 

In a study of $^{12}$C the value of $\beta$ was determined from the first minimum 
of the elastic form factor to be $\beta=1.74$ fm \cite{Bijker2002}. Inspection of Fig.~\ref{splevels} 
shows that for this value of $\beta$, the first 6 neutrons occupy the intrinsic states with 
$\Omega=E_{1/2}^{(+)}$ (black), $E_{3/2}$ (blue) and $E_{1/2}^{(-)}$ (red), so that the last neutron 
in $^{13}$C occupies the intrinsic state with $E_{1/2}^{(-)}$ (red), followed by $E_{1/2}^{(+)}$ (black). 

\begin{table}[h]
\centering
\caption{Resolution of single-particle levels into irreducible representations of $D'_{3h}$. 
Each $E$ level is double degenerate.}
\label{splitting}
\vspace{10pt}
\begin{tabular}{cccc}
\hline
\hline
\noalign{\smallskip}
& $E_{1/2}^{(+)}$ & $E_{1/2}^{(-)}$ & $E_{3/2}$ \\
\noalign{\smallskip}
\hline
\noalign{\smallskip}
$s_{1/2}$ & 1 & 0 & 0 \\
\noalign{\smallskip}
$p_{1/2}$ & 0 & 1 & 0 \\
\noalign{\smallskip}
$p_{3/2}$ & 0 & 1 & 1 \\
\noalign{\smallskip}
$d_{3/2}$ & 1 & 0 & 1 \\
\noalign{\smallskip}
$d_{5/2}$ & 1 & 1 & 1 \\
\noalign{\smallskip}
$f_{5/2}$ & 1 & 1 & 1 \\
\noalign{\smallskip}
$f_{7/2}$ & 2 & 1 & 1 \\
\noalign{\smallskip}
\hline
\hline
\end{tabular}
\end{table}

The representations $E_{1/2}^{(+)}$, $E_{1/2}^{(-)}$ and $E_{3/2}$ can be further 
decomposed into values of $K$ \cite{Bijker2019}.
\ba
\Omega = E_{1/2}^{(+)} &:& K^P = \frac{1}{2}^+, \frac{5}{2}^-, \frac{7}{2}^-, 
\frac{11}{2}^+, \frac{13}{2}^+, \ldots
\nonumber\\
\Omega = E_{1/2}^{(-)} &:& K^P = \frac{1}{2}^-, \frac{5}{2}^+, \frac{7}{2}^+, 
\frac{11}{2}^-, \frac{13}{2}^-, \ldots
\nonumber\\
\Omega = E_{3/2} &:& K^P = \frac{3}{2}^{\pm}, \frac{9}{2}^{\pm}, \frac{15}{2}^{\pm}, \ldots
\label{omega3a}
\ea
with $n=1,2,3,\ldots,$ and $K > 0$. The angular momenta are given by $J=K,K+1,K+2,\ldots$. 
As a result, the rotational sequences for each one of the irreducible representations of $D'_{3h}$ 
are given by (see also Table~\ref{splitting})
\ba
\Omega = E_{1/2}^{(+)} &:& J^P = \frac{1}{2}^+, \frac{3}{2}^+, \frac{5}{2}^{\pm}, 
\frac{7}{2}^+, \left(\frac{7}{2}^-\right)^2, \frac{9}{2}^+, \left(\frac{9}{2}^-\right)^2, \ldots
\nonumber\\
\Omega = E_{1/2}^{(-)} &:& J^P = \frac{1}{2}^-, \frac{3}{2}^-, \frac{5}{2}^{\pm}, 
\left(\frac{7}{2}^+\right)^2, \frac{7}{2}^-, \left(\frac{9}{2}^+\right)^2, \frac{9}{2}^-, \ldots
\nonumber\\
\Omega = E_{3/2} &:& J^P = \frac{3}{2}^{\pm}, \frac{5}{2}^{\pm}, \frac{7}{2}^{\pm}, 
\left(\frac{9}{2}^{\pm}\right)^2, \ldots
\label{omega3b}
\ea
The angular momentum structure of each one of the representations of $D'_{3h}$ is shown in Fig.~\ref{rotbands}.  

\begin{figure}[h]
\centering
\setlength{\unitlength}{0.5pt}
\begin{picture}(790,450)(0,-30)
\thicklines
\put (  0,-30) {\line(1,0){790}}
\put (  0,420) {\line(1,0){790}}
\put (  0,-30) {\line(0,1){450}}
\put (160,-30) {\line(0,1){450}}
\put (390,-30) {\line(0,1){450}}
\put (620,-30) {\line(0,1){450}}
\put (790,-30) {\line(0,1){450}}
\put ( 40,360) {$\bf D_{3h}: A'_1$}
\put (230,360) {$\bf D'_{3h}: E_{1/2}^{(-)}$}
\put (460,360) {$\bf D'_{3h}: E_{1/2}^{(+)}$}
\put (660,360) {$\bf D'_{3h}: E_{3/2}$}
\put ( 30, 60) {\line(1,0){20}}
\put ( 30,120) {\line(1,0){20}}
\put ( 30,260) {\line(1,0){20}}
\put ( 90,180) {\line(1,0){20}}
\put ( 90,260) {\line(1,0){20}}
\put ( 40, 10) {$\bf 0^+$}
\put (100, 10) {$\bf 3^-$}
\put ( 55, 55) {$\bf 0^+$}
\put ( 55,115) {$\bf 2^+$}
\put ( 55,255) {$\bf 4^+$}
\put (115,175) {$\bf 3^-$}
\put (115,255) {$\bf 4^-$}
\put (190, 60) {\line(1,0){20}}
\put (190, 90) {\line(1,0){20}}
\put (190,140) {\line(1,0){20}}
\put (190,210) {\line(1,0){20}}
\put (190,300) {\line(1,0){20}}
\put (250,140) {\line(1,0){20}}
\put (250,210) {\line(1,0){20}}
\put (250,300) {\line(1,0){20}}
\put (310,210) {\line(1,0){20}}
\put (310,300) {\line(1,0){20}}
\put (200, 10) {$\bf \frac{1}{2}^-$}
\put (215, 55) {$\bf \frac{1}{2}^-$}
\put (215, 85) {$\bf \frac{3}{2}^-$}
\put (215,135) {$\bf \frac{5}{2}^-$}
\put (215,205) {$\bf \frac{7}{2}^-$}
\put (215,295) {$\bf \frac{9}{2}^-$}
\put (260, 10) {$\bf \frac{5}{2}^+$}
\put (275,135) {$\bf \frac{5}{2}^+$}
\put (275,205) {$\bf \frac{7}{2}^+$}
\put (275,295) {$\bf \frac{9}{2}^+$}
\put (320, 10) {$\bf \frac{7}{2}^+$}
\put (335,205) {$\bf \frac{7}{2}^+$}
\put (335,295) {$\bf \frac{9}{2}^+$}
\put (420, 60) {\line(1,0){20}}
\put (420, 90) {\line(1,0){20}}
\put (420,140) {\line(1,0){20}}
\put (420,210) {\line(1,0){20}}
\put (420,300) {\line(1,0){20}}
\put (480,140) {\line(1,0){20}}
\put (480,210) {\line(1,0){20}}
\put (480,300) {\line(1,0){20}}
\put (540,210) {\line(1,0){20}}
\put (540,300) {\line(1,0){20}}
\put (430, 10) {$\bf \frac{1}{2}^+$}
\put (445, 55) {$\bf \frac{1}{2}^+$}
\put (445, 85) {$\bf \frac{3}{2}^+$}
\put (445,135) {$\bf \frac{5}{2}^+$}
\put (445,205) {$\bf \frac{7}{2}^+$}
\put (445,295) {$\bf \frac{9}{2}^+$}
\put (490, 10) {$\bf \frac{5}{2}^-$}
\put (505,135) {$\bf \frac{5}{2}^-$}
\put (505,205) {$\bf \frac{7}{2}^-$}
\put (505,295) {$\bf \frac{9}{2}^-$}
\put (550, 10) {$\bf \frac{7}{2}^-$}
\put (565,205) {$\bf \frac{7}{2}^-$}
\put (565,295) {$\bf \frac{9}{2}^-$}
\put (650, 90) {\line(1,0){20}}
\put (650,140) {\line(1,0){20}}
\put (650,210) {\line(1,0){20}}
\put (650,300) {\line(1,0){20}}
\put (710,300) {\line(1,0){20}}
\put (660, 10) {$\bf \frac{3}{2}^\pm$}
\put (675, 85) {$\bf \frac{3}{2}^\pm$}
\put (675,135) {$\bf \frac{5}{2}^\pm$}
\put (675,205) {$\bf \frac{7}{2}^\pm$}
\put (675,295) {$\bf \frac{9}{2}^\pm$}
\put (720, 10) {$\bf \frac{9}{2}^\pm$}
\put (735,295) {$\bf \frac{9}{2}^\pm$}
\end{picture}
\vspace{15pt}
\caption{Structure of rotational bands for a triangular configuration of $\alpha$ particles 
in even-cluster nuclei (first panel) and odd-cluster nuclei with $E_{1/2}^{(-)}$, $E_{1/2}^{(+)}$ 
and $E_{3/2}$ symmetry (second, third and fourth panel). Each rotational band is labeled by the 
quantum numbers $K^P$.}
\label{rotbands}
\end{figure}

The rotational energy spectra can be analyzed with 
\ba
E_{\rm rot}(\Omega,K,J) &=& \varepsilon_{\Omega} + A_{\Omega} 
\left[ J(J+1) + b_{\Omega} K^{2} \right. 
\nonumber\\
&& \hspace{1.5cm} \left. + a_{\Omega} (-1)^{J+1/2}(J+1/2) \delta_{K,1/2} \right] ~,
\label{erot3}
\ea
where $\varepsilon_{\Omega}$ is the intrinsic energy, $A_{\Omega} = \hbar^{2}/2{\cal I}$  
the inertial parameter, $b_{\Omega}$ a Coriolis term, and $a_{\Omega}$ the decoupling parameter. 
The latter term applies only to representations $\Omega = E_{1/2}^{(\pm)}$ and $K^P=1/2^{\pm}$.

\begin{figure}
\centering
\includegraphics[width=4in]{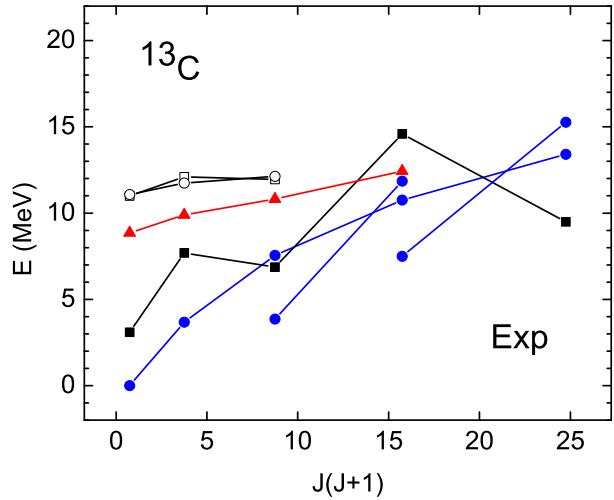}
\caption{Rotational bands in $^{13}$C \cite{Bijker2019}.} 
\label{C13exp}
\end{figure}

Fig.~\ref{C13exp} shows the rotational bands of $^{13}$C. The ground-state band has $K^{P}=1/2^-$ 
and is assigned to the representation $\Omega=E_{1/2}^{(-)}$ of $D_{3h}^{\prime}$ (blue lines and filled 
circles) arising from the coupling of the ground-state band in $^{12}$C to the intrinsic state with 
$E_{1/2}^{(-)}$. According to Eq.~(\ref{omega3a}), this representation contains also $K^{P}=5/2^{+}$ 
and $7/2^{+}$ bands, both of which appear to have been observed. In the shell model, positive parity 
states are expected to occur at much higher energies since they come from the $s$-$d$ shell. 
The first excited rotational band has $K^{P}=1/2^{+}$ which can be assigned to $\Omega=E_{1/2}^{(+)}$ 
(black line and filled squares) arising from the coupling of the ground-state band in $^{12}$C to the 
excited intrinsic state with $E_{1/2}^{(+)}$. In contrast to the ground-state band, this excited band 
has a large decoupling parameter. In addition, Fig.~\ref{C13exp} shows evidence for the occurrence of 
a rotational band at an energy slightly higher than that of the Hoyle state in $^{12}$C which is 
interpreted as the coupling of the Hoyle band in $^{12}$C to the ground-state intrinsic state 
$E_{1/2}^{(-)}$ (red line and filled triangles). 

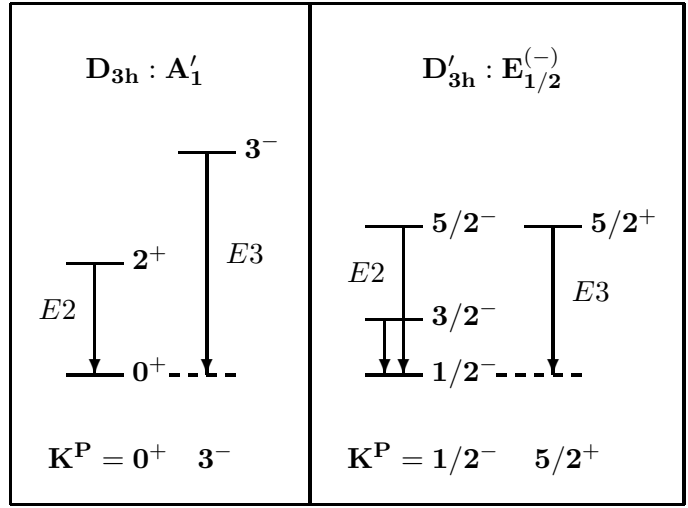
\begin{figure}
\centering
\setlength{\unitlength}{0.7pt}
\begin{picture}(360,270)(0,-10)
\thicklines
\put (  0,-10) {\line(1,0){360}}
\put (  0,260) {\line(1,0){360}}
\put (  0,-10) {\line(0,1){270}}
\put (160,-10) {\line(0,1){270}}
\put (360,-10) {\line(0,1){270}}
\put ( 40,220) {$\bf D_{3h}: A'_1$}
\put (220,220) {$\bf D'_{3h}: E_{1/2}^{(-)}$}
\put ( 30, 60) {\line(1,0){30}}
\put ( 30,120) {\line(1,0){30}}
\put ( 45,120) {\vector(0,-1){60}}
\put ( 15, 90) {$E2$}
\multiput(85,60)(10,0){4}{\line(1,0){5}}
\put ( 90,180) {\line(1,0){30}}
\put (105,180) {\vector(0,-1){120}}
\put (115,120) {$E3$}
\put ( 20, 10) {$\bf K^P=0^+$}
\put (100, 10) {$\bf 3^-$}
\put ( 65, 57) {$\bf 0^+$}
\put ( 65,117) {$\bf 2^+$}
\put (125,177) {$\bf 3^-$}
\put (190, 60) {\line(1,0){30}}
\put (190, 90) {\line(1,0){30}}
\put (190,140) {\line(1,0){30}}
\put (200, 90) {\vector(0,-1){30}}
\put (210,140) {\vector(0,-1){80}}
\put (180,110) {$E2$}
\multiput(260,60)(10,0){5}{\line(1,0){5}}
\put (275,140) {\line(1,0){30}}
\put (290,140) {\vector(0,-1){80}}
\put (300,100) {$E3$}
\put (180, 10) {$\bf K^P=1/2^-$}
\put (225, 57) {$\bf 1/2^-$}
\put (225, 87) {$\bf 3/2^-$}
\put (225,137) {$\bf 5/2^-$}
\put (280, 10) {$\bf 5/2^+$}
\put (310,137) {$\bf 5/2^+$}
\end{picture}
\caption{Electric transitions in $^{12}$C (left) and $^{13}$C (right).}
\label{EL}
\end{figure}

Further evidence for the occurrence of $D_{3h}^{\prime }$ symmetry in $^{13}$C is provided by 
electromagnetic transition rates. The $B(EL)$ values in $^{13}$C are related to those in $^{12}$C 
\cite{Bijker2019} (see also Fig.~\ref{EL}), 
\ba
B(E2;3/2_{1}^{-} \rightarrow 1/2_{1}^{-}) &=& 0.99 \times B(E2;2_{1}^{+}\rightarrow 0_{1}^{+}) ~,
\nonumber\\ 
B(E2;5/2_{1}^{-} \rightarrow 1/2_{1}^{-}) &=& 0.66 \times B(E2;2_{1}^{+}\rightarrow 0_{1}^{+}) ~,
\nonumber\\
B(E3;5/2_{1}^{+} \rightarrow 1/2_{1}^{-}) &=& 0.57 \times B(E3;3_{1}^{-}\rightarrow 0_{1}^{+}) ~.
\ea
The numerical factors arise from combinations of Clebsch-Gordan coefficients. The results 
are shown in Table~\ref{BELC13}. The good agreement with the experimental $B(EL)$ values in $^{13}$C 
shows that the states $3/2_{1}^{-}$, $5/2_{1}^{-}$ and $5/2_{1}^{+}$ belong to the same rotational 
band with $\Omega=E_{1/2}^{(-)}$. Just as in the case of $^{12}$C there is a large octupole transition 
which is obtained in the CSM without the need to introduce effective charges.

\begin{table}[h]
\centering
\caption{$B(EL)$ values in $^{13}$C. Experimental data taken from \cite{ajz91}, 
and the CSM values from \cite{PPNP,Bijker2019}.}
\label{BELC13}
\vspace{10pt}
\begin{tabular}{cccc}
\hline
\noalign{\smallskip}
& CSM & Exp \\ 
\noalign{\smallskip}
\hline
\noalign{\smallskip}
$B(E2;3/2_{1}^{-}\rightarrow 1/2_{1}^{-})$ & $8.3$ & $6.4 \pm 1.5$ & $e^{2}\mbox{fm}^{4}$ \\ 
\noalign{\smallskip}
$B(E2;5/2_{1}^{-}\rightarrow 1/2_{1}^{-})$ & $5.5$ & $5.6 \pm 0.4$ & $e^{2}\mbox{fm}^{4}$ \\ 
\noalign{\smallskip}
$B(E3;5/2_{1}^{+}\rightarrow 1/2_{1}^{-})$ & $42$  & $100 \pm 40$  & $e^{2}\mbox{fm}^{6}$ \\
\noalign{\smallskip}
\hline
\end{tabular}
\end{table}

\section{Summary and conclusions}

In this contribution, we presented a combined analysis of the rotation-vibration spectra and 
electromagnetic transition rates in $^{12}$C and $^{13}$C. A comparison with the theoretical 
predictions of the algebraic cluster model ($^{12}$C) and the cluster shell model ($^{13}$C) 
provides strong evidence for the occurrence of triangular symmetry in these nuclei.  
A characteristic feature of the triangular symmetry is the appearance of rotational 
bands consisting of both positive and negative parity states. The quadrupole and octupole 
transitions in $^{12}$C and $^{13}$C are strongly correlated and only depend on a single 
coefficient $\beta$ whose value was determined independently from the first minimum in the 
elastic form factor of $^{12}$C. The good agreement between theory and experiment supports 
the interpretation of the nucleus $^{13}$C as a system of three $\alpha$-particles in a 
triangular configuration plus an additional neutron moving in the deformed field generated 
by the cluster.

\ack
This work was supported in part by grant IN109017 from DGAPA-UNAM, Mexico.

\end{document}